\documentclass[twocolumn,superscriptaddress,showpacs,prl]{revtex4}
\usepackage{graphicx,epsfig,psfrag}

\newcommand{\be}{\begin{equation}}
\newcommand{\ee}{\end{equation}}
\newcommand{\al}{\alpha}

\newcommand{\lm}{\lambda}
\newcommand{\bea}{\begin{eqnarray}}
\newcommand{\eea}{\end{eqnarray}}

\newcommand{\nn}{\nonumber}

\begin{document}

\noindent
TTP04-09\hfill 
\title{
\boldmath 
Two-Loop Sudakov Form Factor in a Theory with Mass Gap
\unboldmath}
\author{Bernd Feucht}
\affiliation{Bernd Jantzen in later publications}
  \affiliation{Institut f\"ur Theoretische Teilchenphysik,
    Universit\"at Karlsruhe, 76128 Karlsruhe, Germany}
\author{Johann H. K\"uhn}
\affiliation{Institut f\"ur Theoretische Teilchenphysik,
    Universit\"at Karlsruhe, 76128 Karlsruhe, Germany}
\affiliation{Kavli Institute for Theoretical Physics, 
 University of California, Santa Barbara, CA 93106 USA}
\author{Alexander A. Penin}
  \affiliation{Institut f\"ur Theoretische Teilchenphysik,
    Universit\"at Karlsruhe, 76128 Karlsruhe, Germany}
  \affiliation{Institute for Nuclear Research,
    Russian Academy of Sciences, 117312 Moscow, Russia}
\author{Vladimir A. Smirnov}
  \affiliation{Institute for Nuclear Physics,
    Moscow State University, 119992 Moscow, Russia}
\affiliation{II. Institut f{\"u}r Theoretische Physik,
  Universit{\"a}t Hamburg,  22761 Hamburg, Germany}


\begin{abstract}
The two-loop Sudakov form factor is computed in a $U(1)$ model with a massive gauge
boson and a $U(1)\times U(1)$ model with mass gap. We analyze the result 
in the context of  hard and infrared evolution equations and establish
a matching  procedure which relates the  theories with and without 
mass gap setting  the stage for the complete calculation of the  dominant 
two-loop corrections to electroweak processes at high energy.
\end{abstract}

\maketitle

Since the pioneering works by Sudakov \cite{Sud} and Jackiw \cite{Jac} 
the high energy asymptotics of the electromagnetic form factor 
has been the subject of  numerous  investigations. The problem is relevant
for a wide class of phenomenological applications from Drell-Yan
processes to deep inelastic scattering. Recently 
a new wave of interest to the Sudakov asymptotic regime 
has been risen in connection with   higher-order corrections 
to  electroweak processes at high energies
\cite{Kur,Bec1,Bec2,Fad,KPS,CCC,KMPS,HKK,FKM}. 
Experimental and theoretical studies of electroweak interactions have
traditionally explored the range from very low energies, e.g. through
parity violation in atoms, up to energies comparable to the masses of the
$W$- and $Z$-bosons, e.g. at the LEP or the Tevatron.
The advent of multi-TeV colliders like the LHC during the present decade
or a future linear electron-positron
collider will give access to a completely new
energy domain. Once the characteristic energies $\sqrt{s}$ are far larger than the masses of the
$W$- and $Z$-bosons, $M_{W,Z}$, exclusive
reactions like electron-positron (or quark-antiquark) annihilation into a
pair of fermions or gauge bosons will receive  virtual corrections
enhanced by  powers of the  large 
{\it electroweak} logarithm  $\ln\bigl({s/ M_{W,Z}^2}\bigr)$.
The leading double-logarithmic    corrections 
may well amount to ten or even twenty percent in  one-loop
approximation and reach a few percent in two-loop approximation.
Moreover, 
in the TeV region, the subleading logarithms turn out to be equally
important \cite{KPS,KMPS} and a  percent accuracy  of the theoretical estimates 
for the cross sections necessary for the search of new physics 
beyond the standard model can be guaranteed  only by including {\it all} the 
logarithmic two-loop corrections.

The calculation of the two-loop electroweak corrections even in the high
energy limit is an extremely challenging theoretical problem.
It is complicated in particular by  
the presence of the mass gap and mixing  in the
gauge sector.  However, the logarithmic corrections are quite
insensitive to  fine details of the spontaneous symmetry breaking.
The  calculation of the 
leading  logarithmic (LL) electroweak corrections can be performed  
using the fields of the unbroken symmetry phase and how the infrared singular
virtual photon contribution can be separated within the infrared
evolution  equation approach ~\cite{Fad}.  This scheme has been extended
to  the next-to-leading (NLL) and  next-to-next-to-leading 
logarithmic (N$^2$LL) approximation  in   Refs.~\cite{KPS,KMPS}.  

In the  study of Sudakov corrections the analysis of the 
form factor plays a special role since it is the simplest
quantity which includes the 
complete information about the universal {\it collinear} logarithms 
\cite{Fre} directly applicable to a process with an arbitrary
number of fermions.
In this Letter  we formulate a general matching  procedure
which relates the  logarithmic corrections in the 
theories with and without mass gap by combining the hard and infrared 
evolution equation  approach with the explicit two-loop results for 
the form factor in an Abelian gauge model. This 
reduces the calculation of the dominant two-loop  corrections
to   electroweak processes at high energy to a
single-mass problem without mixing.

The structure of the Letter is as follows. First, 
we present the explicit two-loop results for the form factor 
in a $U(1)$ model with a massive gauge boson. Then we introduce 
the  evolution equations, compute the 
two-loop corrections to the form factor in a $U(1)\times U(1)$ model
with  mass gap, and establish the  matching procedure.
Finally we outline how the approach can be applied to 
the calculation of the two-loop electroweak corrections to 
neutral current four-fermion processes.

The vector form factor  ${\cal F}$ determines the
fermion scattering amplitude in an external Abelian field.  
It is a function of the Euclidean momentum transfer
$Q^2=-(p_1-p_2)^2$ where $p_{1,2}$ is the incoming/outgoing fermion
momentum and we consider  on-shell massless fermions, $p_1^2=p_2^2=0$. 
Let us write the perturbative expansion for the form factor 
as  ${\cal F}_\al(M,Q)=\sum_n\left(\al\over 4\pi\right)^nf^{(n)}{\cal F}_B$
where ${\cal F}_B$ corresponds to the Born approximation,  $f^{(0)}=1$.
For the $U(1)$ model with a  gauge boson of  mass $M$
in the Sudakov limit $M/Q\to 0$ the one-loop correction is well known 
\be
f^{(1)}=-{\cal L}^2
+3{\cal L}
-{7\over 2}-{2\over 3}\pi^2\,,
\label{1loopf}
\ee
where ${\cal L}=\ln\left({Q^2/M^2}\right)$ and all the power-suppressed 
terms are neglected.  For the  two-loop term we find
\bea
f^{(2)}&=&{1\over 2}{\cal L}^4-{3}{\cal L}^3
+\left(8+{2\over 3}\pi^2\right){\cal L}^2
-\big(9+4\pi^2
\nn\\
&&
-24\zeta(3)\big){\cal L}
+{25\over 2}+{52\over 3}\pi^2
+80\zeta(3)-{52\over 15}\pi^4
\nn\\
&&
-{32\over 3}\pi^2\ln^22+{32\over 3}\ln^42
+256\,{\rm Li}_4\left({1\over 2}\right)\,,
\label{2loopf}
\eea
where 
$\zeta(3)=1.202057\ldots$ and 
${\rm Li}_4\left({1\over 2}\right)=0.517479\ldots$
are the values of the Riemann's $\zeta$-function
and the polylogarithm, respectively. In Eq.~(\ref{2loopf}) we
do not include  the contribution
due to the virtual fermion loop computed in \cite{FKM}.
For the calculation of the leading power behavior of
the two-loop on-shell vertex diagrams
with two massive propagators in the Sudakov limit
we used the expansion by regions approach
\cite{BenSmi} (for the application to the Sudakov form factor  see
also \cite{KPS}).  The method is based on the factorization of the 
contributions of the dynamical  modes characteristic for  the
Sudakov limit \cite{Ste} in dimensional regularization.
Our result for the  contribution of the hard modes agrees with 
the dimensionally regularized massless result of Ref.~\cite{KraLam}.  
Details of our calculation will be published elsewhere.

\begin{figure}
\psfrag{Q [GeV]}{$Q$ [GeV]}
\epsfig{figure=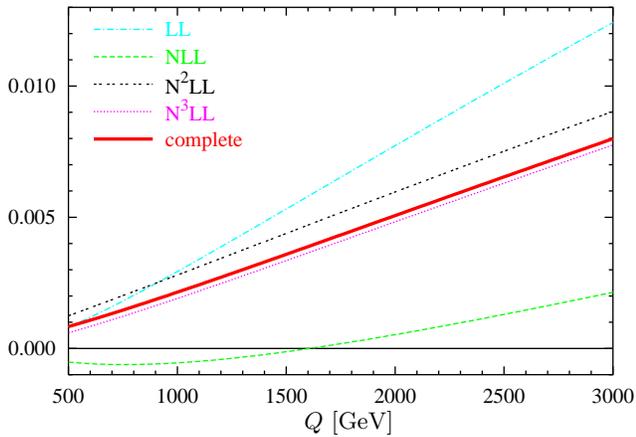,height=6cm}
\caption{\label{fig1} The two-loop correction to the form factor  ${\cal F}_\al(M,Q)$
in LL (including $\al^2{\cal L}^4$), NLL (including $\al^2{\cal L}^3$),
N$^2$LL (including $\al^2{\cal L}^2$),  N$^3$LL (including $\al^2{\cal L}^1$)
approximations and the complete two-loop  correction
as functions of the momentum transfer for $M=80$~GeV, $\al/(4\pi)=3\cdot 10^{-3}$.
}
\end{figure}

\begin{figure}
\psfrag{Q [GeV]}{$Q$ [GeV]}
\epsfig{figure=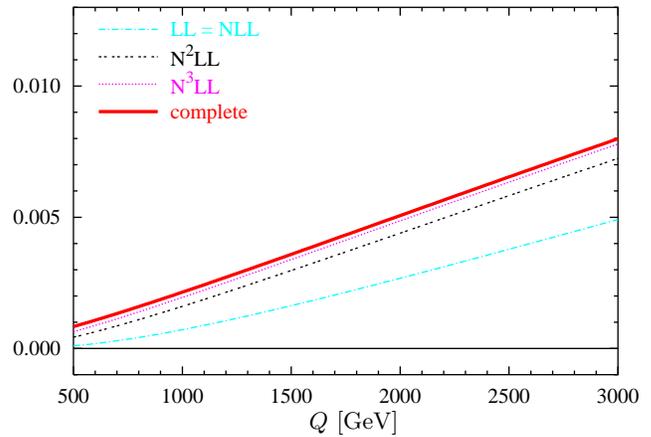,height=6cm}
\caption{\label{fig2} The same as Fig.~\ref{fig1}
after changing the argument of the logarithm. 
}
\end{figure}

In Fig.~\ref{fig1} the numerical results for the two-loop correction  
to the form factor in the different logarithmic approximations
are plotted as  functions of the momentum for 
the values of $M$ and $\al$  typical for electroweak interactions. 
The two-loop  logarithmic terms have a sign-alternating 
structure resulting in significant cancellations.  
In  the region of  a few TeV the form factor does not reach the 
double-logarithmic asymptotics. The quartic, cubic and quadratic  
logarithms are  comparable in magnitude  and  dominate the two-loop corrections. 
Then the logarithmic expansion starts
to converge and, after including the linear-logarithmic contribution, 
provides a very  accurate approximation of the total two-loop correction. 
Such a behavior is typical  for the Sudakov limit and holds 
for the  non-Abelian corrections as well \cite{KMPS,FKM}.
Note that by rescaling $M\to e^{3/4}M$ in the argument of the logarithm 
the  NLL contribution can be made to disappear. That  improves
significantly the
convergence of the logarithmic expansion and prevents the strong
cancellation between the logarithmic terms (see Fig.~\ref{fig2}).  
Still, the N$^3$LL contribution
is a must for the  quantitative approximation.

The asymptotic dependence of the form factor on $Q$ is governed by the 
linear {\it hard} evolution equation  \cite{Mue}. 
As a consequence, the logarithmic corrections exponentiate. 
For the purely Abelian contribution the exponent has a 
particularly simple form 
\bea
{\cal F}_\al(M,Q)&=&\exp\Big\{{\al\over 4\pi}\Big[-{\cal L}^2
+\Big(3-{\al\over 4\pi}\Big(-{3\over 2}+2\pi^2
\nn\\
&&
-24\zeta(3)\Big)+{\cal O}(\al^2)\Big){\cal L}\Big]\Big\}{\cal F}_\al(M,M)\,.
\label{expf}
\eea
The double-logarithmic term 
in the exponent is protected against the  Abelian multiloop  corrections
by  the properties of the light-cone  Wilson loop \cite{KorRad}. 
Our two-loop result determines the next-to-next-to-next-to-next-to-leading
logarithmic (N$^4$LL) approximation of
the form factor  which includes the $\al^n{\cal L}^{m}$  corrections  
with $m=2n-4,\ldots,2n$ to all orders in $\al$.

Let us now turn to the second example with two 
Abelian gauge bosons of the masses $\lm$ and $M$, $\lm\ll M$,
and couplings $\al'$ and  $\al$,  respectively. 
We can  introduce  the {\it infrared} evolution equation which
governs  the dependence of the form factor ${\cal F}(\lm,M,Q)$ 
on $\lm$ \cite{Fad}.   The virtual corrections 
become  divergent in the limit $\lm\to 0$. 
According to the Kinoshita-Lee-Nauenberg theorem \cite{KLN}, these divergences are
cancelled  against the ones of the corrections due to 
the emission of real light gauge bosons of vanishing energy and/or  
collinear to  one of the on-shell fermion lines. 
The singular behavior of the form factor 
must be the same in the full  
$U_{\al'}(1)\times U_{\al}(1)$ theory and the 
effective $U_{\al'}(1)$ model  with only the
light gauge boson.  Thus for $\lm\ll M\ll Q$
the solution of the infrared evolution equation
is given by the exponent of Eq.~(\ref{expf}) with $M$, $\al$ replaced by 
$\lm$, $\al'$, and the form factor can be written in a factorized form 
\be
{\cal F}(\lm,M,Q)=
\tilde{F}(M,Q){\cal F}_{\al'}(\lm,Q)+{\cal O}(\lm/M)\,,
\label{fac}
\ee
where the function $\tilde{F}(M,Q)$ depends on $\al$ and $\al'$ 
and incorporates all the 
logarithms of the form  $\ln\left({Q^2/M^2}\right)$.
It can  be obtained directly by calculating the ratio
\be
\tilde{F}(M,Q)=\left[{{\cal F}(\lm,M,Q)\over
{\cal F}_{\al'}(\lm,Q)}\right]_{\lm\to 0} \,.
\label{tf}
\ee
Since the  function  $\tilde{F}(M,Q)$ does not depend on the
infrared regularization, we compute the ratio in  Eq.~(\ref{tf})
with $\lm=0$ using  dimensional regularization
for the infrared divergences. The method of  calculation of the 
two-loop diagrams with both massive and massless gauge bosons 
is similar to the  purely massive case. We obtain the two-parameter 
perturbative expansion
$\tilde{F}(M,Q)=\sum_{n,m}{\al'{}^n\al^m\over (4\pi)^{n+m}}\tilde
f^{(n,m)}$,
where $\tilde f^{(0,0)}=1$, $\tilde f^{(n,0)}=0$, 
$\tilde f^{(0,m)}=f^{(m)}$, and  the two-loop interference term reads
\bea
\tilde f^{(1,1)}&=&
\big(3-4\pi^2+48\zeta(3)\big){\cal L}
-2+{20\over 3}\pi^2
\nn\\
&&
-84\zeta(3)+{7\over 45}\pi^4\,.
\label{2looptf}
\eea
The numerical structure of the corrections to $\tilde{F}(M,Q)$
is very similar to the one of  ${\cal F}_{\al}(M,Q)$ (see Fig.~\ref{fig1}).

In  the equal mass case,  $\lm=M$, 
we have an additional reparameterization symmetry, and  
the form factor  is determined by Eq.~(\ref{expf})
with  the effective coupling $\bar\al= \al'+\al$ so that  
${\cal F}(M,M,Q)={\cal F}_{\bar\al}(M,Q)$.
We can now write down the matching relation 
\be
{\cal F}(M,M,Q)=
C(M,Q)\tilde{F}(M,Q)
{\cal F}_{\al'}(M,Q)\,,
\label{matf}
\ee
where the  matching coefficient $C(M,Q)$ represents the effect of 
the power-suppressed terms neglected in  Eq.~(\ref{fac}). 
By combining the explicit results for ${\cal F}_{\al'}(M,Q)$ and
$\tilde{F}(M,Q)$  we find the two-loop matching coefficient
\bea
C(M,Q)&=&1+{\al'\al\over(4\pi)^2}
\left[{59\over 4}+{70\over 3}\pi^2+244\zeta(3)-{113\over 15}\pi^4\right.
\nn\\
&&
\left.
-{64\over 3}\pi^2\ln^22
+{64\over 3}\ln^42+512\,{\rm Li}_4\left({1\over 2}
\right)\right]\!.
\label{2loopc}
\eea
Eq.~(\ref{2loopc}) does not contain logarithmic terms,
and up to the N$^3$LL accuracy the product $\tilde{F}(M,Q){\cal F}_{\al'}(\lm,Q)$ 
continuously approaches  ${\cal F}(M,M,Q)$ as $\lm$ goes to
$M$. Therefore, to get  {\it all} the
logarithms of the heavy gauge boson mass
in two-loop approximation for the  theory with mass gap,
it is sufficient to divide the  form factor 
${\cal F}_{\bar\al}(M,Q)$ of the symmetric phase
by the  form factor ${\cal F}_{\al'}(\lm,Q)$ of the effective  
$U_{\al'}(1)$ theory taken at the symmetric point $\lm=M$. 
Thus we have reduced the calculation in the theory with
mass gap to the one in the symmetric theory with a single mass parameter.
Note that the absence of the linear-logarithmic term in Eq.~(\ref{2loopc})
is an exceptional  feature of the Abelian corrections.
The general analysis of the evolution equation \cite{KMPS} 
shows the terms   neglected in  Eq.~(\ref{fac}) to contribute
starting from the N$^3$LL approximation. 
This implies  the absence of the  second and higher powers 
of the logarithm  in the  matching coefficient of  Eq.~(\ref{matf}), 
irrespectively of the gauge group and the mass generation mechanism.  
Moreover, in the approximately equal mass case, $(M-\lm)/M\equiv\delta\ll 1$,
one can compute the form factor as an expansion around the equal mass result.
Up to  N$^2$LL accuracy only the leading term of Eq.~(\ref{fac}) 
contributes and the expansion takes the form 
\bea
\left.{\cal F}(\lm,M,Q)\right|_{\lm\to M}&=&
\left[1-\delta{\al'\over \pi}\left({\cal L}-{3\over 2}\right)
+{\cal O}(\delta^2)\right]
\nn\\
&&\times{\cal F}_{\tilde\al}(M,Q)+{\cal O}(\delta\al'\al{\cal L})\,.
\label{delexp}
\eea
Let us show how the above procedure applies to the 
calculation of two-loop electroweak corrections.
To be specific, we consider a  four-fermion neutral 
current process, which is of primary phenomenological
importance, with light fermions. The four-fermion amplitude can be decomposed into  
(the square of) the form factor and a {\it reduced} amplitude \cite{KPS,KMPS}. 
The latter carries all the Lorentz and isospin indices
and does not contain collinear logarithms in perturbative expansion.
The logarithmic  corrections to the  reduced amplitude
are  obtained  by solving a renormalization group like equation  \cite{Sen}.
The corresponding two-loop anomalous dimensions can be extracted 
from the existing massless QCD  calculations \cite{AGOT}
(see \cite{KMPS,SteTej}). Thus, the problem of the calculation of the 
two-loop  electroweak logarithms in the  four-fermion
processes  reduces  to the analysis of the  form factor.

In Ref.~\cite{KMPS} by analyzing the 
hard  evolution equation 
it has been found that the two-loop electroweak corrections up to 
the  next-to-next-to-leading  (quadratic) logarithms  are not sensitive to 
the  structure of the theory at the electroweak symmetry
breaking scale.  The prediction of Ref.~\cite{KPS,KMPS} for the  two-loop  
logarithmic corrections  fully agrees with the available explicit results
for the light fermion contribution \cite{FKM} and the Abelian contribution
obtained in this Letter. The only trace of the 
Higgs mechanism of the gauge boson mass generation in N$^2$LL
approximation is the  $Z-W$ boson mass splitting which can be
systematically taken into account within an expansion 
around the equal mass approximation similar to Eq.~(\ref{delexp}). 
Thus, the calculation of the two-loop electroweak corrections up to 
the  quadratic logarithms can be performed in two steps
outlined above:
(i) the corrections  are evaluated using the fields of unbroken 
symmetry phase with all the gauge bosons of the same mass $M\approx
M_{Z,W}$ introduced by hand;
(ii) the  QED contribution with an auxiliary photon mass $M$ 
is factorized as in  Eq.~(\ref{matf}) leaving the pure electroweak logarithms. 
The separated  virtual QED corrections accompanied by  the real  photon radiation
in the limit of vanishing photon mass result in the universal infrared safe factor
independent of $M_{Z,W}$.

By contrast, the  N$^3$LL approximation is sensitive to 
fine details of the gauge boson mass generation and 
the coefficient of the linear two-loop electroweak 
logarithm depends {\it e.g.} on the  Higgs boson mass.
For  the full calculation of this coefficient
one has to use  the true mass eigenstates of the  standard model.
Our result, Eqs.~(\ref{2loopf},~\ref{2looptf}), is an example of such a
calculation when applied to the two-loop diagrams with  photon and
$Z$ boson exchanges. We can, however, make  a reasonable approximation
which dramatically simplifies the analysis. Namely,   consider 
a simplified model with a Higgs boson  of zero hypercharge. 
Then the mixing is absent and the hypercharge gauge boson
remains massless.   The interference diagrams including the heavy  $SU_L(2)$ 
and the light  hypercharge $U(1)$  gauge bosons are identical with the  ones of the 
purely Abelian model discussed in this  Letter, where the above two-step procedure
can be applied to get all the two-loop logarithms including the linear term. 
In the standard model the mixing of the gauge bosons 
results in a linear-logarithmic contribution, which is  not accounted 
for within this procedure.  It is, however, suppressed by a small factor 
$\sin^2{\theta_W}\approx 0.2$, with $\theta_W$ being the Weinberg angle.
Therefore,  the above simplified model  gives an estimate of the coefficient 
in front of the linear  electroweak logarithm with $20\%$ accuracy.  
From the  numerical result of Fig.~\ref{fig1}, 
which represents the typical structure of the two-loop 
corrections, we see that a $20\%$ error in this  coefficient  
leads to an uncertainty comparable to the nonlogarithmic 
contribution and is practically negligible. Thus we are able to get an accurate 
estimate of the two-loop correction, which is sufficient 
for  practical applications to the future
collider physics,  by performing the calculations
in the model without mass gap and mixing of the gauge bosons.
The last ingredient necessary to 
complete the calculation of the dominant two-loop electroweak
corrections is the generalization of  Eq.~(\ref{2loopf})
up to the linear-logarithmic term to 
the pure $SU_L(2)$ gauge model with the Higgs mechanism 
of mass generation.   Note that up to the quadratic logarithm 
the two-loop corrections are predicted by the 
evolution equation \cite{KMPS}.

To conclude, we have obtained the complete  results for the  
two-loop corrections to the vector form factor in the Sudakov limit 
in   Abelian  theories with one massive gauge boson or with two 
gauge bosons of essentially different masses. The results are in 
full agreement with the predictions of the evolution equation approach.
We have formulated a systematic procedure 
of factorizing  the infrared singular virtual corrections 
and reducing the calculation in the theory with mass gap to the
single-mass problem. 
The analysis can directly be generalized to the   
standard model  with  spontaneous breaking of the $SU_L(2)\times U(1)$ 
theory to  low-energy QED.
This  solves the principal problems of the calculation of the 
dominant two-loop  electroweak corrections 
to the neutral current four-fermion processes 
which are mandatory for the high-precision physics at 
the LHC and the next generation of linear colliders.

\begin{acknowledgments}
We thank S. Pozzorini for useful comments  on the manuscript.
J.H.K.  acknowledges the hospitality of Kavli Institute for
Theoretical Physics and partial support by the NSF under
Grant No. PHY99-0794.
The work of J.H.K and A.A.P.  was supported in part by BMBF Grant No.\
05HT4VKA/3 and Sonderforschungsbereich Transregio 9. The work of V.A.S. was
supported in part by Volkswagen Foundation
Contract No. I/77788, and DFG Mercator Visiting Professorship No. Ha
202/110-1.
\end{acknowledgments}

\end{document}